# Intrinsic Radiation in Lutetium Based PET Detector: Advantages and Disadvantages

Qingyang Wei (魏清阳)[1,2]

1. Department of Electrical Engineering, Tsinghua University, Beijing 100084, China
2. Key Laboratory of Particle & Radiation Imaging, Ministry of Education, Beijing, 100084, China

**Abstract**: Lutetium (Lu) based scintillators such as LSO and LYSO, are widely used in modern PET detectors due to their high stopping power for 511 keV gamma rays, high light yield and short decay time. However, 2.6% of naturally occurring Lu is $^{176}$Lu, a long-lived radioactive element including a beta decay and three major simultaneous gamma decays. This phenomenon introduces random events to PET systems that affects the system performance. On the other hand, the advantages of intrinsic radiation of $^{176}$Lu (IRL) continues to be exploited. In this paper, research literatures about IRL in PET detectors are reviewed. Details about the adverse effects of IRL to PET and their solutions, as well as the useful applications are presented and discussed.

## 1 PET and PET detector

Positron emission tomography (PET) is a nuclear medicine and functional imaging technique [1]. The system detects pairs of 511keV gamma rays emitted indirectly by positron-emitting radionuclides ($^{11}$C, $^{13}$N, $^{15}$O, $^{18}$F, etc.), which are labeled on biologically active molecules and introduced into the bodies. Three-dimensional images of tracer concentration within the bodies are then mathematically reconstructed. In clinical examinations, PET is a powerful tool for early diagnosing cancers, heart diseases and brain diseases, et al [1]. And PET is also applied in pre-clinical small animal imaging for drug development, pathological studies and gene expression studies, et al [3].

PET machines are mainly built from scintillator detectors consisting of inorganic scintillators and optical photon detectors (PMT, APD or SiPM)[4]. The gamma rays deposit energy in the scintillators and generate optical photons, which are detected by the optical photon detectors. The scintillator of the early PET detector is high light yield NaI(Tl) [6]. Whereas, it has disadvantages for PET including low stopping power, long

The paper has been submitted to Chinese Physics C
Supported by China Postdoctoral Science Foundation (2014M550745), National Natural Science Foundation of China (No. 11375096), National Natural Science Foundation of China (11275105), and Tsinghua University Initiative Scientific Research Program (20131089289).
Email: weiqingyang@tsinghua.edu.cn

decay time and hydroscopic property. In 1970s, BGO with high stopping power was discovered and substituted NaI(Tl) for PET. However, BGO is still not ideal for PET due to its low light yield and long decay time [5]. In the late 1980s, lutetium oxyorthosilicate (LSO) were discovered by Melcher *et al* [7] and then other lutetium based scintillators (LBS) such as LYSO and LFS were also found. With properties of high stopping power, high light yield and fast decay time, LBS is quite suitable for PET especially time of flight PET (TOF-PET)[8], thus LBS becomes the best choice of modern PET detectors. The parameters of some scintillators used in PET detectors are listed in table1.

Table 1 Properties of some scintillators used in PET detectors. Note that the parameters have slight variation due to the change of the dopants in the scintillator growth [5].

|  | NaI(Tl) | BGO | LSO | LYSO | LFS |
|---|---|---|---|---|---|
| Linear attenuation coeff.(cm$^{-1}$) | 0.34 | 0.92 | 0.87 | 0.86 | 0.82 |
| Light yield (%NaI(Tl)) | 100 | 15 | 75 | 80 | 77 |
| Decay constant (ns) | 230 | 300 | 40 | 41 | 35 |

## 2 Intrinsic radiation of Lutetium-176 (IRL)

The naturally occurring Lutetium consists a fraction of $^{176}$Lu (abundance 2.6%), a long-lived radioactive element with a half-life ($T_{1/2}$) of ~$3.6\times10^{10}$ year [9]. Figure 1 shows the decay scheme of $^{176}$Lu. The main process of $^{176}$Lu $\rightarrow$ $^{176}$Hf decay is a β-particle emission (maximum energy 596 keV) and three simultaneous γ-ray emissions (energies: 88, 202, and 307 keV). The count rate ($R$) of a milliliter scintillator can be calculated with equation 1:

$$R = \frac{n\rho N_A}{M} \times 2.6\% \times (1 - e^{-0.693/T_{1/2}}) \approx \frac{n\rho N_A}{M} \times 2.6\% \times \frac{0.693}{T_{1/2}} \quad , \tag{1}$$

Where, $n$ is the subscript number of Lu in the scintillator molecular formula, $\rho$ is the density, $N_A$ is the Avogadro constant ($6.02\times10^{23}$), and $M$ is the molar mass.

Take LSO (Lu$_2$SiO$_5$) for example, calculated as equation 2, $R_{LSO}$ equals to 307 Bq/ml.

$$R_{LSO} = \frac{2 \times 7.4 g/ml \times 6.02 \times 10^{23}}{460} \times 2.6\% \times \frac{0.693}{3.6\times10^{10} \times 365 \times 24 \times 3600} = 307 Bq/ml, \tag{2}$$

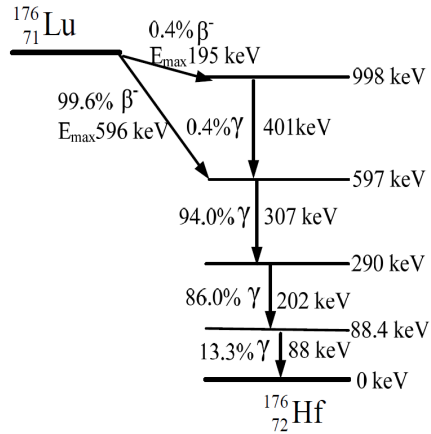

Figure 1 Decay scheme of $^{176}$Lu[10]

In a PET detector, the β particles emitted from the LBS array, given the short range, mostly deposit energies in the same LBS array. While, γ-rays can be detected not only in the same LBS array, but also in other LBS arrays by emitting outside the first LBS array. Therefore, there are different types of events including single photon events, double photon events and multiple photon events. The decay particles of a $^{176}$Lu element only detected in a same LBS array is a single photon event (SPE) (Fig. 2(a)). An escaped γ-particle detected by another LBS array is a true intrinsic coincidence event (TICE) (Fig. 2(b)). Two single events occurring in a coincidence time window (CTW) is a random intrinsic coincidence event (RICE) (Fig. 2 (c)). More than one γ-particle escaping and detected by different LBS arrays is a multiple photon event. A true intrinsic coincidence event and a single event generated in a CTW is also identified as a multiple photon event (Fig2.d).

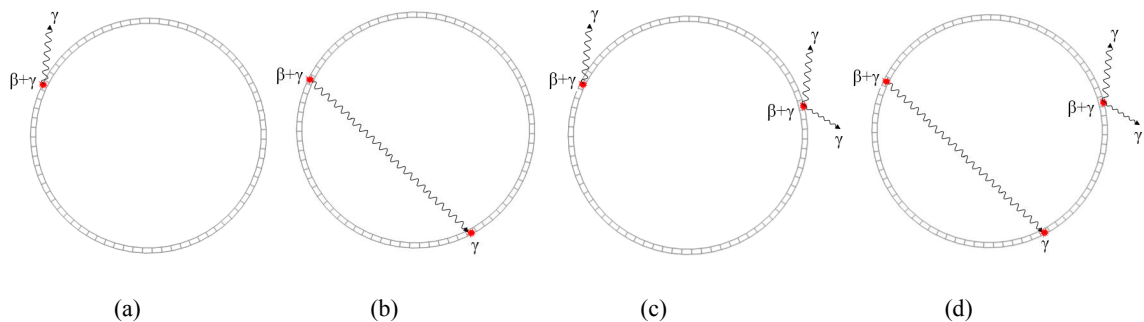

(a)          (b)          (c)          (d)

Figure 2 Several typical IRL events in a PET system. (a) a single-photon event, (b) a double-photon event (true intrinsic coincidence), (c) a double-photon event (random intrinsic coincidence), (d) a multiple-photon event.

The count rate of SPE is proportional to the amount of the scintillator being used. The count rate of TICE is determined by the mount of the scintillator, the size of crystal and the geometry of the system. And the count rate of RICE is proportional to the square of the mount of scintillator and the value of CTW.

Figure 3 shows the GATE-simulated [11] spectrums of IRL in a small animal PET (InliView 3000 [13]). As shown in Fig. 3(a), the energy spectrum of the SPE in PET detector is a broad beta spectrum distorted by discrete shifts in energy caused by the simultaneous detection of one, two or three of the gamma photons. The count rate of TICE is ~5 kcps. Fig. 3(b) shows the energy spectrum of the TICE without TOF. The peaks of 202 and 307 keV are more obvious than the peaks in SPE. Fig. 3(c) shows the energy spectrum of the TICE with TOF. The first trigger and the second trigger can be distinguished where the second triggers are the escaping gamma photons. Thus the photopeaks of 202 and 308 keV are easiest distinguished.

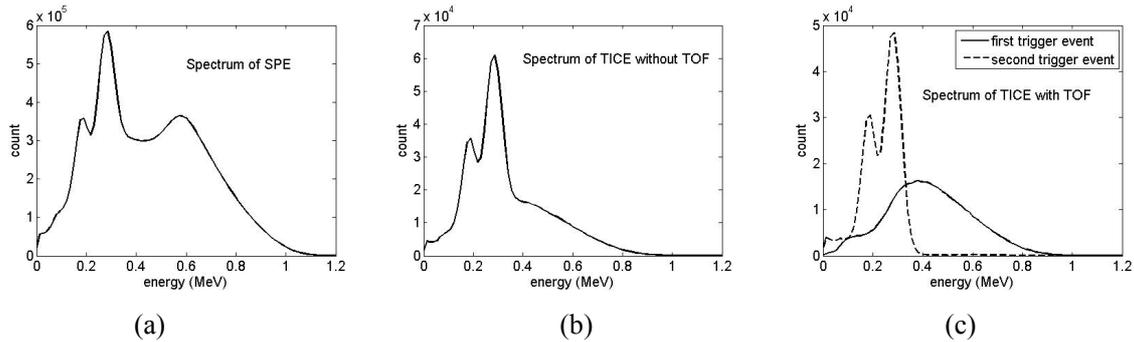

Figure 3 The simulated energy spectrums of Lu background radiation in a small animal PET (InliView 3000), energy resolution 20%@511keV. (a) the spectrum of SPE. (b) the spectrum of the TICE without TOF information. (c) the spectrum of the TICE with TOF.

## 3 The drawbacks of IRL in PET

The effects IRL in PET imaging have been well documented. It is usually not a big concern in most *in-vivo* situations. However, IRL becomes problematic for some special studies such as low activity imaging, long axial FOV PET, and PET/SPECT imaging.

### 3.1 common PET imaging

The IRL affects the system performance by generating additional prompt and random coincidences. Because the SPE energy spectrum of the $^{176}$Lu is a wider distribution covering the photopeak of 511keV, the intrinsic random coincidences could not be prevented completely. Although, the intrinsic random coincidences can be corrected by measuring the delayed coincidence or using some other correction methods [14], they alters the timing performance of the system prolonging the dead time of the detectors [15]. The intrinsic true coincidences will introduce blank noise to the reconstructed image and the scatter fraction will be overestimated with NEMA protocol without considering the Lu intrinsic true coincidences [16].

The count rates of intrinsic random coincidences and intrinsic true coincidences are both depended on the energy window. Usually, the energy window is recommended as 350-650keV in a Lu-based clinical PET system. Under this energy window, most intrinsic true coincidence events can be prevented because most of the prompt gamma of Lu background radiation is 202 and 307 keV which below 350 keV. Moreover, the Lu background radiation is usually very low relative to the injected activity in clinical applications [19], therefore, IRL is not sufficient to affect routine clinical PET scanning.

Compared to the clinical systems, preclinical small animal PETs have much lower injected activities, poorer energy resolution due to the small crystal and less scatter effect in small animals. To improve the sensitivity, the energy window is usually recommended as 250keV-750 keV. Thus, the impact of IRL on the performance of small animal PET systems must not be disregarded [19].

The effects of IRL on PET systems will be further reduced, but not eliminated entirely, as newer generations of detectors and electronics permit the use of more narrow energy acceptance windows and coincidence time windows.

**3.2 low activity PET imaging**

There are some applications of PET imaging done with a low activity level (the injection activity below 1 kBq for some situations), for instance, cell-trafficking studies [20], gene expression imaging [21] or in-beam PET imaging [22]. In these cases, the IRL may seriously affect the imaging performance with the conventional energy window [23]. The minimum detectable activity value is increased due to the Lu background radiation. An available method to reduce the number of Lu intrinsic random coincidence events is using a narrow energy window and the TOF information. Therefore, it requires the Lu based system with a good energy resolution and a good TOF information. In 2014, Yoshida [23] proposed a novel method by combing TOF and multiple coincidence information to further reduce the effects of IRL.

**3.3 Long axial FOV PET**

The axial field of view (AFOV) of current clinical whole-body PET scanners range from 15–22 cm which have limited sensitivity. Recently, a long AFOV (2m) PET was proposed [25] which could image the whole-body at one time and would significantly increase the sensitivity. It would open the door to new applications, such as whole-body parametric imaging of pharmacological kinetics and systemic imaging of radiolabeled stem cell progenitor cell populations [25]. The single event count rate of the IRL is proportional to the amount of the scintillator being used that means it will 10 times of the common PET system in a 2m long AFOV PET. Thus, there will be very high random events in the long AFOV PET. To reduce the impact of random events, variable coincidence time windows for LORs should be employed [25].

**3.4 PET/SPECT imaging**

Animal PET scanners with inserted collimators were developed to do animal SPECT imaging [26]. Due to the low efficiency of SPECT imaging and the energy range of SPECT imaging overlapping with IRL energy spectrum, IRL may have serious impacts on SPECT imaging. The solution method is subtracting the contribution of Lu background by doing a long time of background scan [27]. The pattern of Lu background radiation can be reduced with the correction. However, the minimal detectable target activity will be higher for the Lu-based SPECT system as compared to typical SPECT systems which do not have crystal intrinsic activity [27].

## 4 The benefits of IRL in PET

Everything has two sides. The IRL is an interest spot for researchers and has been utilized in PET in many directions.

### 4.1 For detector module developing

The IRL can be used as an approximate flood source to assist PET detector module design instead of using additional sources. The background radiation is fast enough to generate a flood histogram of a block detector in several minutes [28]. It can help to quickly test the block design of crystal surface processing and reflect schemes. In additional, the Lu-based scintillator array can also be used as a multiple gamma-rays source for other detectors to do energy calibration or energy nonlinear response measurement.

The IRL can also help the study of depth of interaction PET detector (DOI-PET). For example, the scintillators' internal Lu-176 beta decay was applied to measure DOI functions for a dual end readout DOI - PET [29].

### 4.2 For PET quality assurance (QA)

PET requires a common QA procedure to ensure the system working with a good condition. The IRL is able to do a fast test for daily or weekly QA. Fig. 4 shows the 2Dmaps of the Lu coincidence event, where Fig. 4(a) is a relative normal one, while Fig. 4(b) shows that a sector of the system has some problems which need to be fixed.

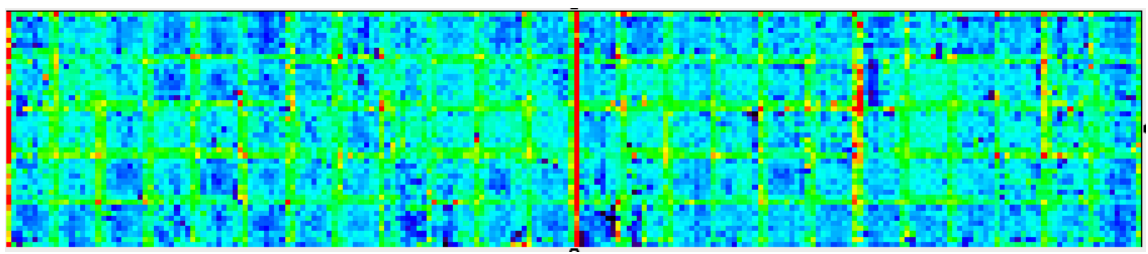

(a)

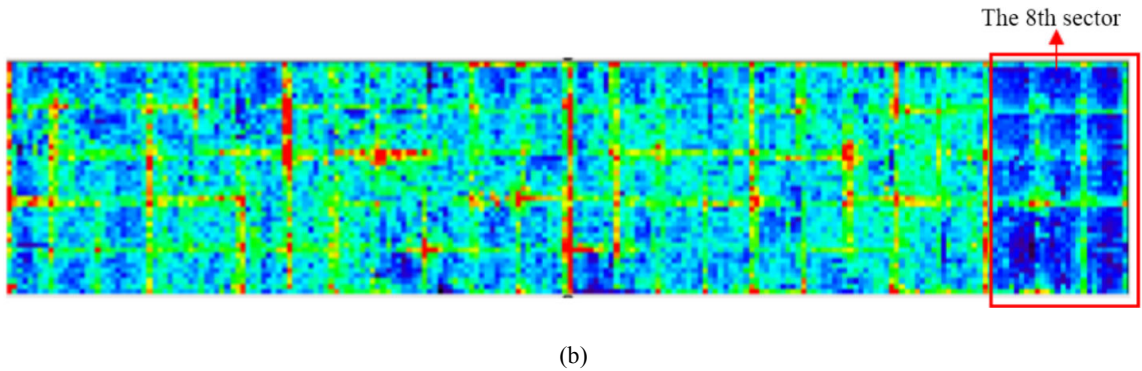

(b)

Figure 4 2Dmaps of the Lu background coincidence events of the InliView-3000 small PET system. (a) A relative normal 2Dmap, some blocks have bad performance and (b) is an un-normal 2Dmap indicated that a problem occurred in the eighth sector.

**4.3 For monitoring PET channel gain drift**

It is necessary to do energy calibration for PET detectors in order to identify the energy of each detector unit. Moreover, the calibration should be repeated over time due to the condition variation of the photon detectors, such as the PMT aging [30] and the temperature changing of APD or SiPM. The IRL was applied as a system self-calibration means to monitor the random or systematic changes or drift in the system [30]. The method consists of tracking the position of the energy peak 597 keV (sum of all three gamma rays), it was proved that the Lu based method obtained a calibration functionally equivalent to the standard calibration method with external 511-keV sources.

**4.4 For Time Alignment of TOF-PET**

Time alignment of a TOF-PET system is an important calibration procedure as the nature of the data collected in PET is inherently with variable time shifts. There are many methods to obtain the time corrections for the crystals in the PET scanner [31]. All of which require a radioactive source, except the method proposed by Rothfuss which is based on IRL [32]. The time of flight is measured between the two crystals involved in a Lu true intrinsic coincidence event. A time of flight can also be calculated between any two crystals with positional information acquired from the fixed geometry of PET scanner. The time correction for the crystal pair can then be equated by taking the measured time difference and subtracting the geometrical calculated time of flight. It was found the timing alignment performed with IRL resulted in performance as good as the method using a uniform gamma source phantom [32].

**4.5 For generating transmission image**

Lu intrinsic true coincidence events are feasibility to perform as a transmission source to generate transmission image of the scan object in PET. Furthermore, in a TOF-PET system, with information of the TOF and the chord length between two crystals in coincidence, a second time window can be set to observe transmission events simultaneously to positron emission events. Thus, the transmission and emission image can be reconstructed simultaneously. It is observed that the flux of the background activity is high enough to create useful transmission images with acquisition time of 10 min [33]. This method can provide an attenuation solution to the system that do not have or require a coupled CT scanner, such as PET-MR machines or dedicated PET machines with specific applications such as brain imaging. The transmission image can also be employed to perform scatter corrections or starting images for algorithms that estimate emission and attenuation simultaneously [33].

**4.6 For system geometrical calibration**

The transmission image of Lu intrinsic coincidence events could also be applied for system geometrical calibration. We developed a small animal PET/SPECT/CT system (InliView 3000) consisting of a LYSO based PET scanner, a collimator SPECT by sharing the PET scanner and a cone-beam CT based on CMOS detector. The IRL was utilized to image the collimator to do geometrical calibration. Fig. 5 shows the transmission image of the collimator. Moreover, we designed a multi-tungsten-alloy-ball phantom to be imaged by IRL and CT scanner. The two types of transmission images are used to generate the transformation matrix between PET and CT system. Our proposed method is expected to replace the traditional multi-point or multi-line gamma sources which make the geometrical calibration more convenient.

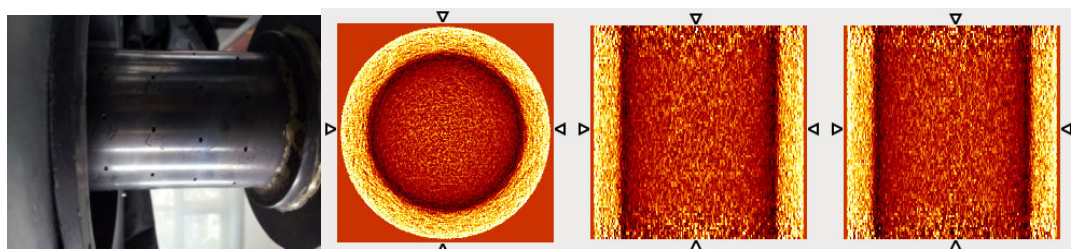

Figure 5 The inserted multi-pinhole collimator in the InliView-3000 (left), the trasaxial, views of the collimator reconstructed image using IRL data (right)[34]

# 5 Conclusions

Lutetium based scintillators with good properties are the first choice of PET detectors. The IRL affects the performance of the PET systems. With a narrow energy window and a narrow coincidence time window, the IRL can be ignored for conventional clinical PET studies while it cannot be disregarded for small animal PETs. For some special systems and applications, such as low activity imaging, long axial FOV PET and PET detector based SPECT imaging, the impact of IRL will be significant.

On the other hand, IRL becomes a powerful tool. The existing of IRL can be utilized to assist detector design, to do system quality assurance, to monitor channel gain drift, to perform time alignment of TOF-PET, to generate transmission image and to do geometrical calibration, et al.